\documentclass{aa}
\usepackage{amsmath,amssymb,bm}
\usepackage{graphicx}
\usepackage[varg]{txfonts}                      % A&A style
\usepackage{flushend}
\usepackage{appendix}
\usepackage[colorlinks, citecolor=blue]{hyperref}

\newcommand{\Bep}{B_\mathrm{ep}}
\newcommand{\Bet}{B_\mathrm{et}}

\begin{document}

\title{Helicity shedding by flux rope ejection} 

\author{B. Kliem    %\inst{1}
   \and N. Seehafer %\inst{1}
       }

\institute{Institute of Physics and Astronomy, University of Potsdam, 
           Potsdam 14476, Germany}

% \titlerunning{Helicity shedding by flux rope ejection}
% \authorrunning{B. Kliem \& N. Seehafer}

\date{Received 12 October 2021 / Accepted 13 November 2021} 

\abstract
% Context: 
{}
% Aims: 
{We quantitatively address the conjecture that magnetic helicity must be shed from the Sun by eruptions launching coronal mass ejections in order to limit its accumulation in each hemisphere.} 
% Methods: 
{By varying the ratio of guide and strapping field and the flux rope twist in a parametric simulation study of flux rope ejection from approximately marginally 
stable force-free equilibria, different ratios of self- and mutual helicity are set and the onset of the torus or helical kink instability is obtained.} 
% Results: 
{The helicity shed is found to vary over a broad range from a minor to a major part of the initial helicity, with self helicity being largely or completely shed and mutual helicity, which makes up the larger part of the initial helicity, being shed only partly. Torus-unstable configurations with subcritical twist and without a guide field shed up to about two-thirds of the initial helicity, while a highly twisted, kink-unstable configuration sheds only about one-quarter. The parametric study also yields stable force-free flux rope equilibria up to a total flux-normalized helicity of 0.25, with a ratio of self- to total helicity of 0.32 and a ratio of flux rope to external poloidal flux of 0.94.} 
% Conclusions: 
{These results numerically demonstrate the conjecture of helicity shedding by coronal mass ejections and provide a first account of its parametric dependence. Both self- and mutual helicity are shed significantly; this reduces the total initial helicity by a fraction of $\sim\!0.4\mbox{--}0.65$ for typical source region parameters.} 
% 250 words


\maketitle

\section{Introduction}
\label{s:intro}

The solar dynamo generates flux with predominantly positive (right-handed) helicity in the southern hemisphere and predominantly negative (left-handed) helicity in the northern hemisphere \citep{Hale1925, Seehafer1990, Pevtsov&al1995}. New flux from the dynamo region emerges primarily at large, active-region scales into the atmosphere, where it undergoes a cascade to small scales driven by the motions in the convection zone \cite[e.g.,][]{Parnell&al2009}. However, the simultaneous cascade of the helicity is thought to proceed primarily toward large scales, although a weaker cascade toward small scales has also been found \citep{Alexakis&al2006}. While small-scale flux elements can submerge and re-emerge with the convection cells in the upper convection zone, the steep density stratification in the photosphere presents a barrier to the submergence of flux at the larger scales, with the dividing line estimated to lie at roughly 1~Mm \citep{vanBallegooijen&Martens1989}. As a result, positive (negative) helicity is thought to accumulate in the large-scale coronal field in the southern (northern) hemisphere. 

The Sun does have options to prevent the helicity from accumulating indefinitely, but two of them proceed very slowly, on the timescale of the magnetic cycle, because they are driven by the meridional flow. One is the annihilation of helicity when flux reconnects across the equatorial plane, which is facilitated by the equatorward migration of the active-region belts in the course of the solar cycle. The other is the transport of flux toward the adjacent pole, there joining the open flux, which can shed the helicity through torsional Alfv\'en waves propagating away from the Sun. As these processes operate on such a long timescale, the helicity would accumulate in each hemisphere to very high levels in the course of a cycle if there were not an option to shed the helicity much faster. 

For the corona, whose flux is rooted in the photosphere and partly extends into the interplanetary space, the relative magnetic helicity is the relevant, gauge-invariant quantity to be considered \citep{Berger&Field1984}. As the relative magnetic helicity is carried by the electric currents in the volume, which also carry the free magnetic energy, the accumulation of helicity is a plausible driver of the coronal field toward instability. Therefore, it has been conjectured that coronal mass ejections (CMEs) are the relevant process of helicity shedding from the corona \citep{Rust1994, LowBC1996}. CMEs presumably result from an instability of the coronal field on active-region scales \citep{vanTend&Kuperus1978, Forbes&Isenberg1991, Kliem&Torok2006}; they occur at least once in every major active region, as well as in the quiet Sun, and carry away a significant fraction of the source region flux in the form of a large-scale flux rope \citep{QiuJ&al2007}. 

The conjecture of helicity shedding by CMEs is observationally supported \citep{Demoulin&al2002, Green&al2002} and widely accepted, although, to our knowledge, the effectiveness of erupting flux ropes at shedding helicity has not yet been investigated quantitatively. 
A preliminary simulation study of helicity shedding by a highly twisted, kink-unstable flux rope yielded a rather low effectiveness: only approximately one-fifth of the initial helicity was shed \citep{Kliem&al2011}. As many active regions experience only one CME in their lifetime, this effectiveness might be too small to prevent an unrealistic accumulation of helicity. 

Here, we perform the first systematic investigation of helicity shedding by simulating the eruption and ejection of flux ropes susceptible to the far more relevant torus instability, and comparing this with the helicity shedding by a kink-unstable flux rope and by the flux rope in a data-constrained model of a specific solar event. Specifically, we focus on the dependence of the shedding on the initial partition into self- and mutual helicity.

\section{Numerical methods}
\label{s:methods}

We model CMEs as ideal magnetohydrodynamic (MHD) instability of a force-free flux rope \citep{Sakurai1976, vanTend&Kuperus1978, Kliem&Torok2006}. The one-fluid ideal MHD equations are integrated numerically in the zero-beta limit, neglecting any thermodynamic effects, which are of only secondary importance for the helicity shedding, and neglecting gravity. Ohmic resistivity is also neglected to approach the almost ideal MHD conditions in the corona as closely as possible. Magnetic reconnection, required for a CME \citep{LinJ&Forbes2000}, is enabled by numerical diffusion in the vertical current sheet that steepens as the unstable flux rope rises. The resulting reconnection rate has been found to lie quite close (within a factor 2) to the observationally estimated one for two carefully modeled solar eruptions \citep{Kliem&al2013, Hassanin&Kliem2016}, although the magnetic diffusion is only numerical and exceeds the coronal magnetic diffusion by many orders of magnitude. The equations are given in \citet{Torok&Kliem2003} for example, where the adopted numerical scheme is also detailed. 

Our modified Lax-Wendroff scheme replaces the numerically stabilizing but highly diffusive Lax step by artificial diffusion \citep{Sato&Hayashi1979} which is of similar structure. The coefficient of the artificial diffusion is tailored for each simulation as a function of time, and if necessary also as a function of space, resulting in a reduced numerical diffusion of the magnetic field compared to the Lax term by a factor $10^3$. For numerical stability, the scheme also includes a small kinematic viscosity. 

As the initial condition, we employ the analytical, approximate force-free equilibrium of a toroidal current channel and flux rope (the tokamak equilibrium) in the specific form suggested by \citet{Titov&Demoulin1999} as a basic model for solar active regions. The current channel of major radius $R$ and minor radius $a$ is situated in the vertical ($y$-$z$) plane, with the center of the torus at $\bm{r}=(0,0,-d)$. The stabilizing external poloidal (``strapping'') field is provided by a pair of magnetic point sources, $\pm q$, located at $\bm{r}=(\pm L,0,-d)$, respectively, whose resulting flux concentrations in the photosphere, $\{z\!=\!0\}$, at $x\approx\pm L$ model a bipolar solar active region. An external toroidal (``guide" or ``shear'') field component is introduced by adding an antiparallel pair of dipoles under the footpoints of the flux rope. Their depth is $\approx2.7R$, determined such that the field line passing through the footpoints of the flux rope axis closely follows the axis in the corona. The model for the initial density is derived from the initial field through $\rho_0(\bm{x})=B_0(\bm{x})^{1.5}$, a choice implying a slow upward and sideward decrease of the initial Alfv\'en velocity from the core of the model active region, as on the Sun, and found to enable the modeling of CMEs in close quantitative agreement with observations \cite[e.g.,][]{Torok&Kliem2005, Kliem&al2012, Kliem&al2013}. The initial velocity is set to zero, $\bm{u}_0=0$. 

This configuration is discretized on a Cartesian grid with a closed bottom boundary at $z=0$, closed side boundaries at $x=\pm l_x$ and $y=l_y$, and an open upper boundary at $z=l_z$. Point-symmetric mirroring about the $z$ axis of the variables in the boundary at $y=0$ takes advantage of the symmetry in the configuration which is preserved throughout the evolution. Photospheric motions are negligible during the short timescales of solar eruptions, and so we set $\bm{u}(x,y,0,t)=0$, which implies that the vertical photospheric field, $B_z(x,y,0)$, is conserved. The side boundaries are chosen to be closed for numerical convenience, as open boundaries can introduce numerical artifacts, especially if inflows develop; this requires the boundaries to be placed sufficiently far away such that the erupting flux never reaches them. Here $\bm{u}=0$ in the boundary, which preserves the normal field component and the density in the boundary, while the tangential field components are allowed to vary to minimize the influence of the boundary on the rising flux rope. The fluid advects freely across the upper boundary; this is implemented by using extrapolation of the velocity onto the ghost layers of the grid. 

The MHD variables are normalized in the natural way \cite[e.g.,][]{Torok&Kliem2003} by using the apex height of the geometric flux rope axis, $h_\mathrm{a}=R-d$, as the length unit, the initial Alfv\'en velocity (and field strength; density) at the apex point of the flux rope axis, $V_\mathrm{A0}=B_0(0,0,h_\mathrm{a})/(\mu_0\rho_0(0,0,h_\mathrm{a}))^{1/2}$, as the velocity (and field; density) unit, and the resulting Alfv\'en time, $\tau_\mathrm{A}=h_\mathrm{a}/V_\mathrm{A0}$, as the unit of time. A core set of configurations uses a uniform geometry of the current channel, given by $d=0.1$ and $a=0.6$. To evaluate the influence of the geometry, two further sets, one with varying $d$, the other with varying $a$, are also included. For each configuration, $L$ is numerically determined such that the initial configuration is close to marginal stability (slightly supercritical) with respect to the torus instability, with the lower and upper bounds on the critical value, $L_\mathrm{cr}$, differing by less than 5\%. For the kink-unstable comparison run, a smaller $a$ yields a supercritical twist and $L$ is chosen to be moderately subcritical. For each set of these parameters, the source strength, $q$, of the strapping field is given by the analytical equilibrium condition. The guide field strength is a free parameter and varied from run to run, as detailed in Section~\ref{ss:parametric}. The grid of $251\times180\times240$ points for the half cube is stretched outward from the origin to permit a large box size, $l_x=l_y=32$ and $l_z=40$, while the central volume, which fully includes the initial flux rope, is well resolved, $\Delta_{x,y,z}=0.04$. The stretching also very efficiently damps any outward traveling perturbation (e.g., that from the initial relaxation of the analytical equilibrium), so that we do not observe any reflection of perturbations at the side boundaries.

\section{Computation of helicities}
\label{s:computation}

We calculate the relative magnetic helicity \citep{Berger&Field1984,Finn&Antonsen1985} in
the rectangular box $V=\{x,y,z\,\mbox{:}\,-l_x\le x\le l_x,\, -l_y\le y\le l_y,\, 0\le z\le l_z\}$
according to 
\begin{align}
H=&\int_{\text{Box volume}} \vec{A}\cdot\vec{B}\,\mathrm{d}^3\vec{r} \nonumber \\
&+\int_{\text{Side faces}}\phi_{\mathrm{C}}\vec{A}_{\mathrm{C}} \, \mathrm{d}\vec{S}
+\int_{\text{Top face}} ( \vec{A} \times \vec{A}_{\mathrm{C}}) \, \mathrm{d}\vec{S} \,,
\label{e:helicity}
\end{align}
where $\vec{A}$ and $\vec{A}_{\mathrm{C}}$ are special vector potentials for $\vec{B}$ and for
the current-free field with the same normal component at the box 
boundary, $\vec{B}_{\mathrm{C}}$, and $\phi_{\mathrm{C}}$ is a scalar potential for $\vec{B}_{\mathrm{C}}$, that is
\begin{equation}
\vec{B}=\nabla\times\vec{A} \,, \;
\vec{B}_{\mathrm{C}} = \nabla\times\vec{A}_{\mathrm{C}}=- \nabla\phi_{\mathrm{C}} \,, 
\label{e:B_C=gradient}
\end{equation}
\begin{equation}
\Delta\phi_{\mathrm{C}}=0 \,, \;
\left.\nabla\phi_{\mathrm{C}}\cdot\vec{\hat{n}}\right|_{\partial V} = \left.-B_n\right|_{\partial V} \,,
\end{equation}
where $\vec{\hat{n}}$ is the exterior unit normal at the boundary $\partial V$ of $V$.
The two vector potentials are given by
\begin{equation}
\vec{A}= \vec{A}_{\mathrm{C}}(x,y,0)-\vec{\hat{z}}\times
 \int_0^z \vec{B}(x,y,z')\, \mathrm{d} z'
\label{e:vector_potential_A}
\end{equation}
and
\begin{equation}
\vec{A}_\mathrm{C}=\vec{\hat{z}}\times\nabla \left(\int_0^z \phi_\mathrm{C}(x,y,z')\, \mathrm{d}z'
 +g_\mathrm{C}(x,y)\right) \,,
\label{e:vector_potential_A_C}
\end{equation}
where $\vec{\hat{z}}$ is the unit vector in the vertical ($z$) direction
and $g_\mathrm{C}(x,y)$ is a solution of the two-dimensional Poisson equation
\begin{equation}
 \left( \partial_x^2 + \partial_y^2 \right) g_\mathrm{C}(x,y)=B_z(x,y,0) \,.
 \label{e:2D_Poisson}
\end{equation}

Our procedure for calculating $H$ is an extension of that derived by \citet{DeVore2000} for the half space $z\ge 0$, where $H$ is given by the first line in Eq.~(\ref{e:helicity}). 
The adaption of the method of \citeauthor{DeVore2000}, characterized in particular by the gauge
$\vec{A}\cdot\vec{\hat{z}} = \vec{A}_\mathrm{C}\cdot\vec{\hat{z}} = 0$ of the vector potentials, to the case of a finite rectangular box was first proposed by \citet{Valori&al2012}; Eqs.~(\ref{e:helicity})--(\ref{e:2D_Poisson}) are easily obtained from their expressions 
(see Appendix \ref{appendix:equations}).
Other algorithms for calculating the relative magnetic helicity in a rectangular box as a functional 
of the field $\vec{B}$ in the box  were suggested, for instance, by \citet{Rudenko&Myshyakov2011},
\citet{Thalmann&al2011},  and \citet{Yang&al2013a,Yang&al2018}. A review of helicity calculation methods for finite volumes was given by \citet{Valori&al2016}.

Defining the closed field in $V$,
\begin{equation}
 \vec{B}_{\mathrm{cl}}=\vec{B}-\vec{B}_{\mathrm{C}},
 \label{e:B_cl}
\end{equation}
and its vector potential, $\vec{A}_{\mathrm{cl}}=\vec{A}-\vec{A}_{\mathrm{C}}$, the helicity can be decomposed into self helicity of the closed field,  
\begin{equation}
 H_\mathrm{self}=\int_V \vec{A}_{\mathrm{cl}}\cdot\vec{B}_{\mathrm{cl}}\, \mathrm{d}^3\vec{r,}
 \label{e:H_self}
\end{equation}
and mutual helicity between the closed and open potential field, 
\begin{equation}
 H_\mathrm{mutual}=2 \int_V \vec{A}_{\mathrm{C}}\cdot\vec{B}_{\mathrm{cl}}\, \mathrm{d}^3\vec{r;}
 \label{e:H_mutual}
\end{equation}
see \citet{Berger1999}. The relation $H=H_\mathrm{self}+H_\mathrm{mutual}$ is satisfied by our numerically computed helicities to better than 0.3\%. The closed field $\vec{B}_{\mathrm{cl}}$ is identical to the current-carrying part of the field, and therefore the self helicity is also referred to as the current-carrying helicity.

\section{Helicity shedding} 
\label{s:shedding}

\subsection{Reference run: no guide field}
\label{ss:reference}

First, we describe the evolution of a torus-unstable configuration with vanishing guide field, which sheds helicity quite effectively and is taken to be our reference (Run~T1). For $d=0.1$ and $a=0.6$, marginal stability with respect to the torus instability is given for a ``sunspot distance'' slightly above the value of $L=1.3$, which is chosen for this run to initiate the eruption. The threshold of the torus instability is given in terms of the decay index of the external poloidal field, $n(z)=-d\ln \Bep(z)/d\ln z$, at the apex point of the flux rope's magnetic axis, which is slightly offset from the geometric axis, $h_\mathrm{m}=1.1$. Experience with many different systems has shown that the threshold value lies in the range $n_\mathrm{cr}\sim1\dots2$ and depends on various parameters of the equilibrium (in ways yet to be determined). The canonical value of $n_\mathrm{cr}\sim3/2$ \citep{Bateman1978} appears to provide a reasonable representative value for approximately semicircular flux ropes in the absence of a guide field \citep{Torok&Kliem2007, Aulanier&al2010, FanY2010, Zuccarello&al2015}. For $L=1.3,$ we have $n(z\!=\!h_\mathrm{m})=1.29$, which is slightly above the numerically determined threshold for the given configuration and is also close to the value of the full analytical expression for $n(R)$ in \citet{Kliem&Torok2006}. The twist, averaged over the cross-section of the current channel, is $|\Phi|=2.45\pi$, and sufficiently small for stability against the helical kink mode \citep{Hood&Priest1981, Torok&al2004}. 

%~~~~~~~~~~~~~~~~~~~~~~~~~~~~~~~~~~~~~~~~~~~~~~~~~~~~~~~~~~~~~~~~~~~~~~~~~~
\begin{figure}  % [h]                                                Fig. 1
\centering
\resizebox{.95\hsize}{!}{\includegraphics{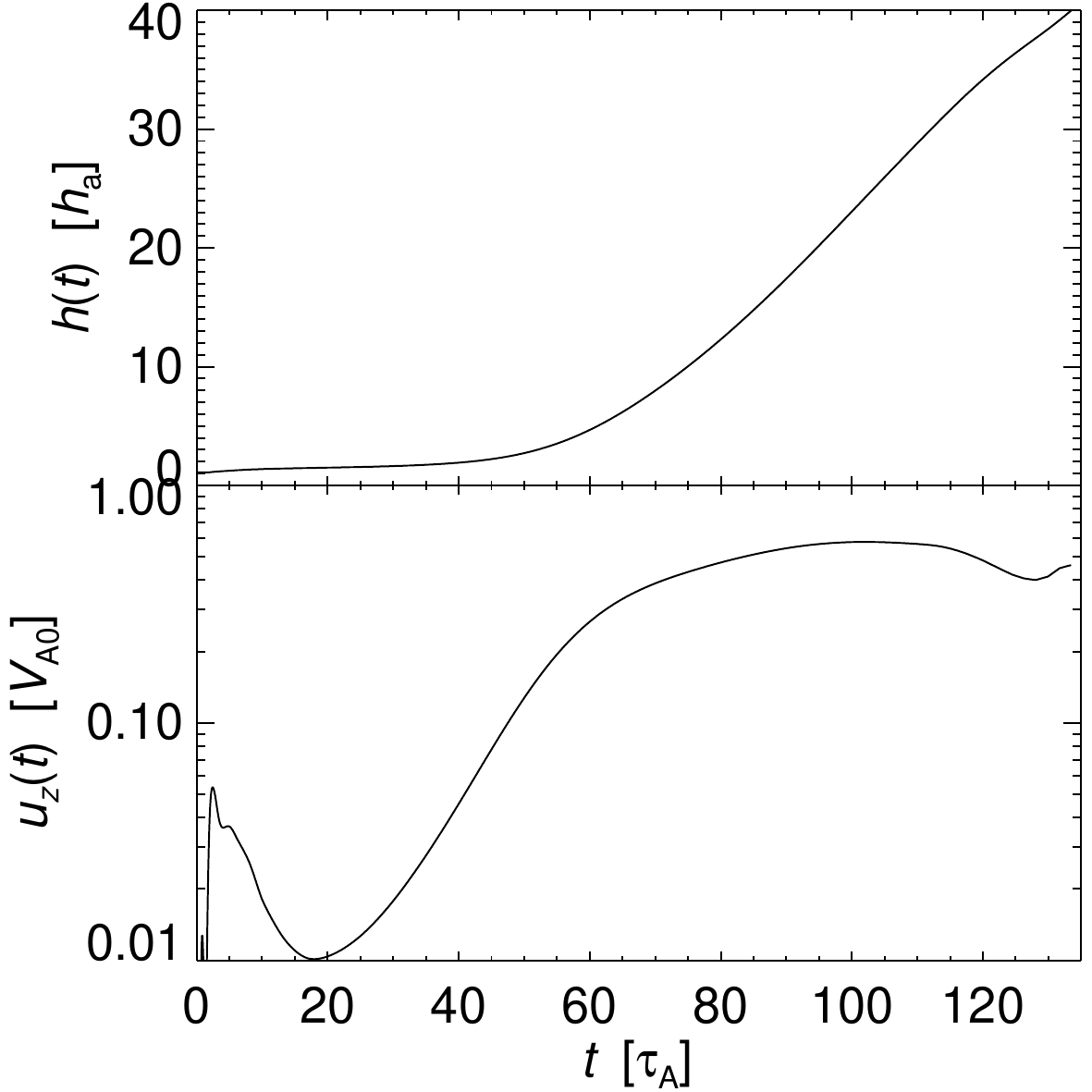}}  % {f1.eps}}  % <-- A&A LaTeX style
\caption[]{Rise profile of the flux rope apex in the reference run T1, showing height, $h(t)$, and velocity, $\log u_z(t)$.} 
\label{f:rise}
\end{figure}
%~~~~~~~~~~~~~~~~~~~~~~~~~~~~~~~~~~~~~~~~~~~~~~~~~~~~~~~~~~~~~~~~~~~~~~~~~~

%~~~~~~~~~~~~~~~~~~~~~~~~~~~~~~~~~~~~~~~~~~~~~~~~~~~~~~~~~~~~~~~~~~~~~~~~~~
\begin{figure*}  % [h]                                               Fig. 2
\centering
\resizebox{.95\hsize}{!}{\includegraphics{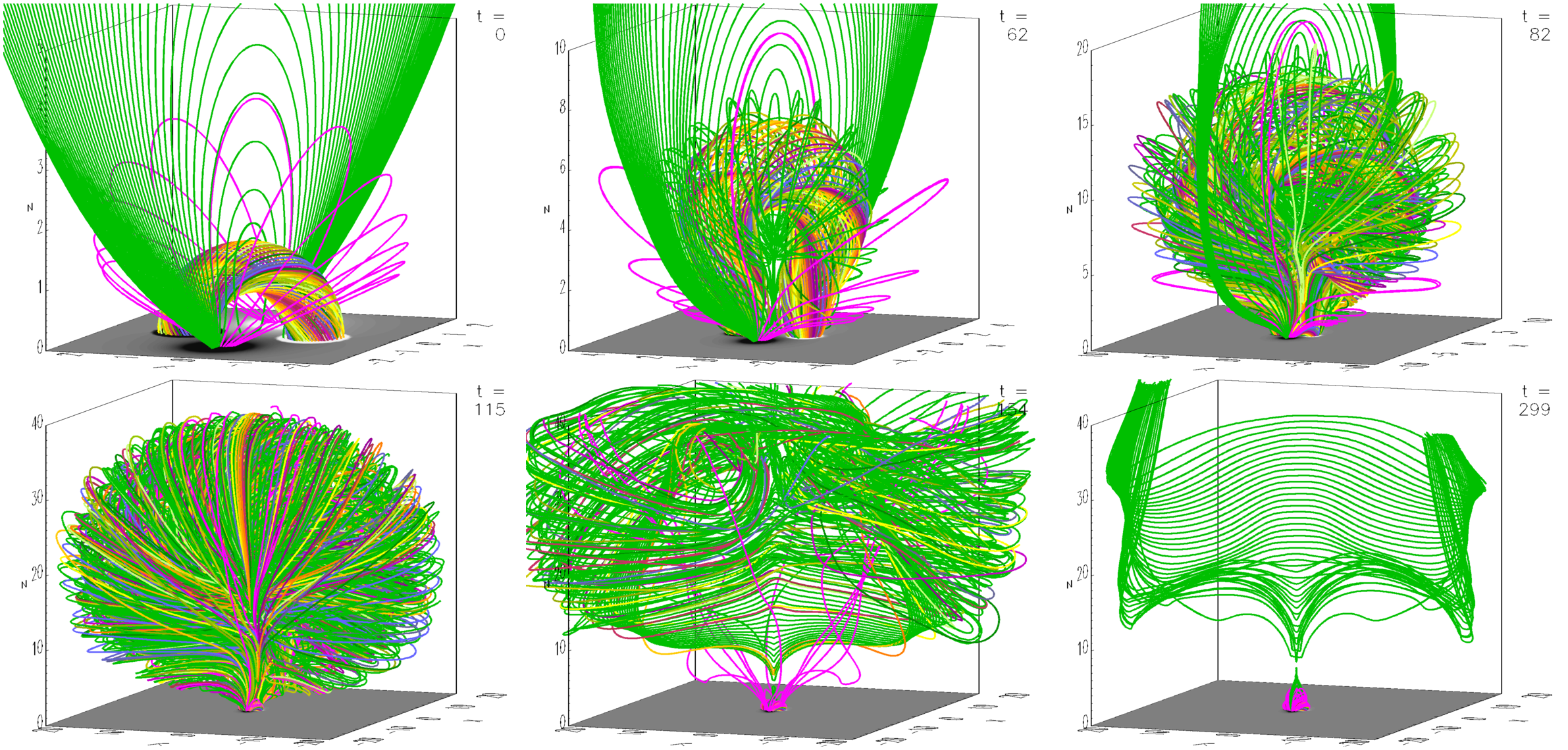}}  % {f2.eps}}
\caption[]{Snapshots of the unstable flux rope in Run~T1 visualized by rainbow-colored field lines. Green field lines are started at the $z$-axis (with uniform spacing) to visualize the overlying flux and the vertical current sheet that forms under the rising flux rope. Pink field lines visualize the ambient flux in the vicinity of the flux rope, which recloses after the eruption. The  magnetogram, $B_z(x,y,0)$, is shown in gray scale; a logarithmic scaling is used because the sunspot flux is rather concentrated due to the small value of $d$.} 
\label{f:fieldlines}
\end{figure*}
%~~~~~~~~~~~~~~~~~~~~~~~~~~~~~~~~~~~~~~~~~~~~~~~~~~~~~~~~~~~~~~~~~~~~~~~~~~

%~~~~~~~~~~~~~~~~~~~~~~~~~~~~~~~~~~~~~~~~~~~~~~~~~~~~~~~~~~~~~~~~~~~~~~~~~~
\begin{figure*}  % [h]                                               Fig. 3
\centering
\resizebox{.95\hsize}{!}{\includegraphics{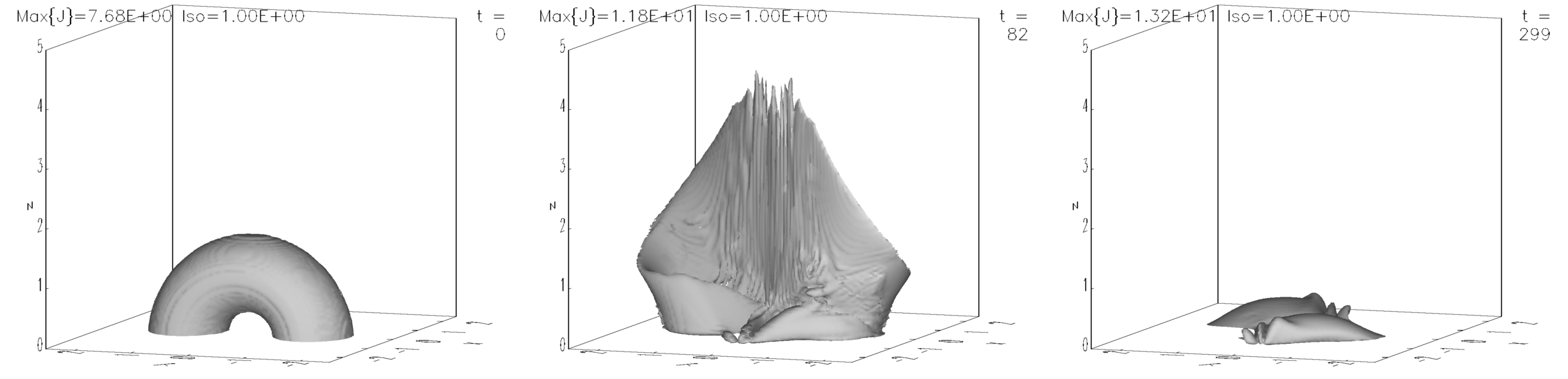}}  % {f3.eps}}
\caption[]{Isosurfaces of current density, $J$, for Run~T1 showing, from left to right:
 the initial flux rope; 
 the vertical current sheet during the evolution of the instability (also wrapping around the bottom part of the flux rope legs in the form of a sigmoid); 
 and the currents remaining after the eruption (which are already visible at $t=82$).} 
\label{f:current}
\end{figure*}
%~~~~~~~~~~~~~~~~~~~~~~~~~~~~~~~~~~~~~~~~~~~~~~~~~~~~~~~~~~~~~~~~~~~~~~~~~~

Figure~\ref{f:rise} shows the rise profile of the fluid element at the apex point of the flux rope axis. The run starts with a phase of numerical relaxation that attempts to reach a numerical equilibrium from the initial approximate analytical equilibrium. The torus instability develops out of this equilibrium from $t\approx18$, initiated by the velocities that develop during the relaxation and act as a small perturbation. The occurrence of instability is clear from the initially nearly perfect exponential rise up to $t\approx50$. The flux rope reaches a rise velocity of $u_z\approx0.6$, and its apex point leaves the box at $t\approx133$. Figures~\ref{f:fieldlines} and \ref{f:current} show snapshots of the flux rope, the initial current channel, the vertical current sheet spawned by the eruption, and the remaining volume currents. The latter are concentrated in the area of the two prominent flux bundles in the re-closed field of the model active region. These remain from the initial flux rope after the rope has reconnected with ambient flux in the vertical current sheet and connect the flux rope footpoints with the opposite-polarity sunspots (see the snapshots at $t\ge82$ in Figure~\ref{f:fieldlines}), which is close to the potential field \cite[see, e.g., Figure~6(e) in][]{Hassanin&Kliem2016}. Such reconnection has been observed in many eruptions of a Titov-D\'emoulin flux rope \cite[e.g.,][]{Torok&Kliem2005} and other models of solar eruptions \citep{Aulanier&Dudik2019} and certainly occurs on the Sun, although the frequency of its occurrence is yet to be determined. 

Figure~\ref{f:helicity} shows the relative magnetic helicity in the box (Eq.~(\ref{e:helicity})), normalized by the unsigned flux in the magnetogram area, $F=(F_+-F_-)/2$, versus\ time for this run. The helicity begins to decrease rapidly when the upper edge of the current-carrying flux rope reaches the upper boundary at $t\approx127$, and levels off after the top part of the flux rope has nearly completely left the box at $t\approx200$. A major part (64\%) of the initial helicity is shed by $t\approx300$ (the very slow subsequent increase in the helicity is considered to be due to the development of weak current density ripples in the central-difference numerical integration and is disregarded). 

The initial helicity, $H_0$, in this configuration is partitioned into self- and mutual helicity as $H_\mathrm{self,0}=0.33H_0$ and $H_\mathrm{mutual,0}=0.67H_0$. The initial self helicity due to the twist in the initial flux rope completely leaves the system with the flux of the CME bubble (Figure~\ref{f:fieldlines}); there is even a slight overshoot (buildup of a small opposite self helicity) in this run, which is not essential in the total helicity budget. Of the initial mutual helicity, 41\% is shed by the eruption, and the rest is carried by the volume currents in and around the two flux bundles that remain from the flux rope (Figure~\ref{f:current} at $t=299$). 

The computation of the helicity was checked by extending the integrals in Equation~(\ref{e:helicity}) to a lower part of the box only, $z=0\mbox{--}10$. The initial helicity was found to be identical to the one in Figure~\ref{f:helicity} (to within $1\times10^{-3}$); that is, in the top part of the box, the computed potential field is nearly identical to the ambient field of the Titov-D\'emoulin flux rope, so that the contribution to the total relative helicity is negligible. The helicity versus\ time was found to follow the curve in Figure~\ref{f:helicity} up to the arrival of the top part of the flux rope at the virtual boundary at $t\approx73$ and to show a slightly more rapid decrease by the same amount, $\Delta H/F^2=0.13$, by $t\approx155$, i.e., the same shedding effectiveness.

\subsection{Parametric study}
\label{ss:parametric}

\subsubsection{Effect of the guide field}
\label{sss:guide}

%~~~~~~~~~~~~~~~~~~~~~~~~~~~~~~~~~~~~~~~~~~~~~~~~~~~~~~~~~~~~~~~~~~~~~~~~~~
\begin{figure}  % [h]                                                Fig. 4
\centering
\resizebox{.95\hsize}{!}{\includegraphics{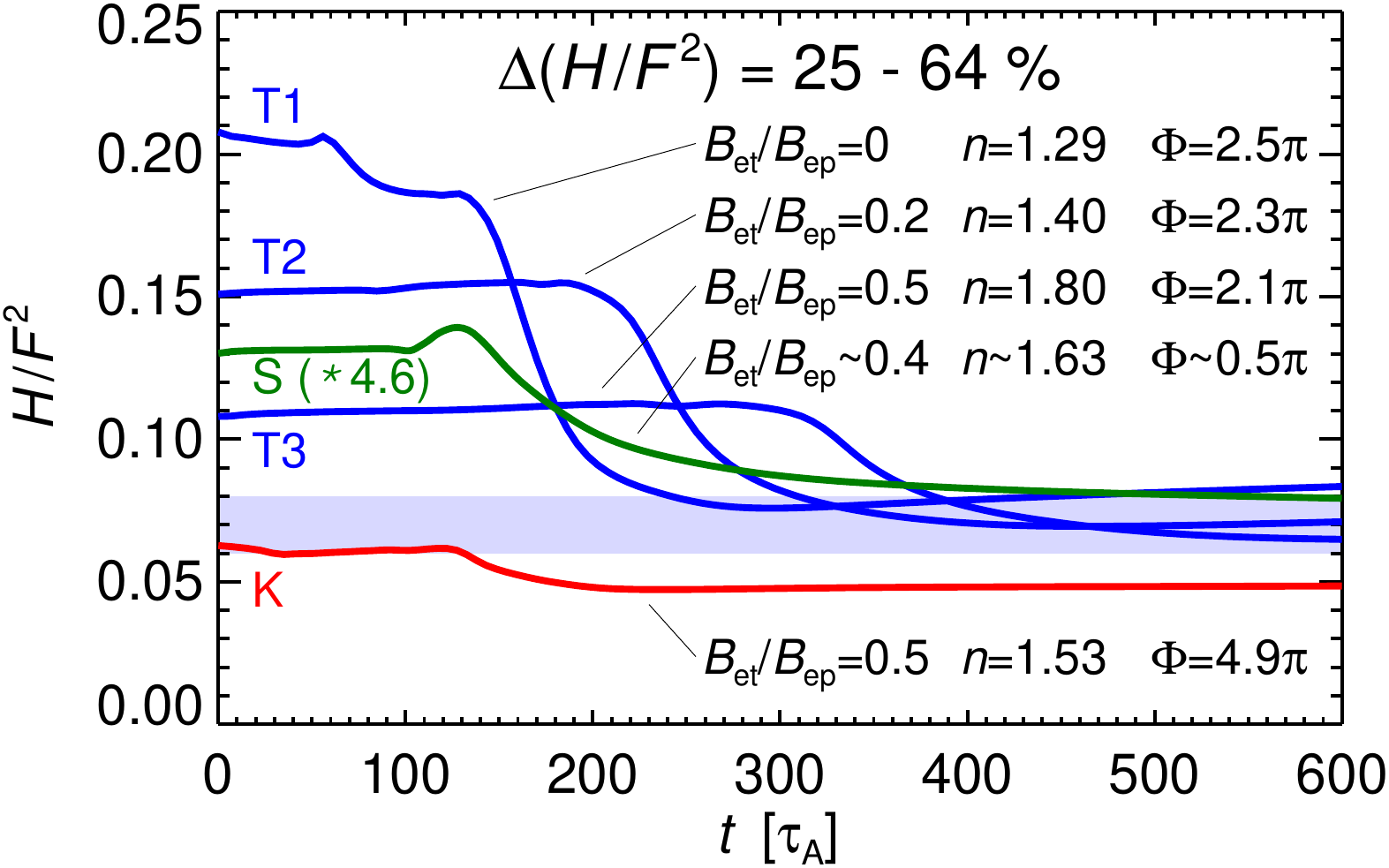}}  % {f4.eps}}
\caption[]{Flux-normalized relative magnetic helicity in the box vs.\ time for the key 
runs analyzed in this paper (absolute values are shown). The initial helicity of Run~S is scaled, placing it in the middle between $H_0/F^2$ for Runs~T2 and T3 based on the similarity of the parameters. The length unit and resulting Alfv\'en time of Run~S are rescaled from the values in \citet{Kliem&al2013} to conform to the choice for the other runs here (the apex height of $0.1R_\odot$ at the onset of instability is used as length unit). The curves are labeled with the respective values of the ratio of external toroidal (guide) and external poloidal (strapping) field, $\Bet/\Bep$, decay index, $n$, at the initial flux rope apex, and twist averaged over the cross-section of the current channel, $\Phi$. The range of the conjectured base level of the mutual helicity in the configurations T1--T3, $\approx\!0.06\mbox{--}0.08$, is shaded in light blue.} 
\label{f:helicity}
\end{figure}
%~~~~~~~~~~~~~~~~~~~~~~~~~~~~~~~~~~~~~~~~~~~~~~~~~~~~~~~~~~~~~~~~~~~~~~~~~~

%*******************************************************************************
\begin{table*}[t]                                                       % Tab. 1
\caption{Parameter values of the Runs T0--T4, K, and S.}
\centering 
\begin{tabular}{llllllll}
\hline\hline 
Run                            & T0              & T1              & T2              & T3             & T4              & K               & S           \\
\hline 
$\Bet/\Bep$                    & 0               & 0               & 0.2             & 0.5            & 0.5             & 0.5             & $\sim\!0.4$ \\
$d$                            & 0.1             & 0.1             & 0.1             & 0.1            & 0.375           & 0.1             & $\dots$     \\
$a$                            & 0.9             & 0.6             & 0.6             & 0.6            & 0.6             & 0.27            & $\dots$     \\
$L$                            & 1.3             & 1.3             & 1.2             & 0.9            & 0.65            & 1.0             & $\dots$     \\
$n(z\!=\!h_\mathrm{m})$        & 1.45            & 1.29            & 1.40            & 1.80           & 1.73            & 1.53            & 1.5--1.75   \\

$\Phi/\pi$                     & 1.7             & 2.5             & 2.3             & 2.1            & 2.4             & 4.9             & $\sim\!0.5$ \\
$H_0/F^2$                      & 0.29            & 0.21            & 0.15            & 0.11           & 0.10            & 0.063           & 0.028       \\
$H_\mathrm{self,0}/F^2$        & 0.084  (~~29\%) & 0.068  (~~33\%) & 0.044  (~~29\%) & 0.022 (~~20\%) & 0.020  (~~19\%) & 0.010  (~~16\%) & 0.008  (28\%) \tablefootmark{a}\\
$H_\mathrm{mutual,0}/F^2$      & 0.21~~ (~~71\%) & 0.14~~ (~~67\%) & 0.11~~ (~~71\%) & 0.086 (~~80\%) & 0.084  (~~81\%) & 0.053  (~~84\%) & 0.021  (72\%) \tablefootmark{a}\\
$\Delta H/F^2$                 & 0.20~~ (~~70\%) & 0.13~~ (~~64\%) & 0.081  (~~54\%) & 0.043 (~~39\%) & 0.040  (~~38\%) & 0.015  (~~24\%) & 0.012~~ (42\%) \tablefootmark{b}\\
$\Delta H_\mathrm{self}/F^2$   & 0.084   (100\%) & 0.075   (100\%) & 0.044  (100\%)  & 0.022  (100\%) & 0.021   (100\%) & 0.012   (100\%) & 0.007  (87\%) \tablefootmark{b}\\
$\Delta H_\mathrm{mutual}/F^2$ & 0.118  (~~57\%) & 0.057  (~~41\%) & 0.037  (~~34\%) & 0.021 (~~24\%) & 0.019  (~~23\%) & 0.003 (~~~~6\%) & 0.005  (25\%) \tablefootmark{b}\\
\hline
\end{tabular}
\tablefoot{\tablefoottext{a}{Percentage of $H_0/F^2$ is given in parentheses.}
           \tablefoottext{b}{Shedding effectiveness is given in parentheses as a percentage of the initial value.}}
\label{t:shedding}
\end{table*}
%*******************************************************************************

The strong difference in the effectiveness of helicity shedding between the torus-unstable configuration of Section~\ref{ss:reference} and the kink-unstable configuration in \citet{Kliem&al2011} motivates us to study the dependence of the shedding on the partition into mutual- and self helicity of the initial configuration. The twist and corresponding self helicity of the flux rope play at most a secondary role in the dynamics of the former case, but are the driver of the initial instability in the latter. First we study the effect of the partition on purely torus-unstable configurations, which are most likely relevant for the majority of CMEs. To this end, we take advantage of the fact that the strength of the external toroidal field component in the initial equilibrium can be varied freely without significantly changing the residual forces. In two further runs, we keep the geometry from Run~T1 and set the strength of the external toroidal field such that its ratio to the equilibrium value of the external poloidal field at the apex of the flux rope axis is $\Bet/\Bep=0.2$ (Run~T2) and $\Bet/\Bep=0.5$ (Run~T3). As the external toroidal field has a stabilizing effect on the torus and helical kink instabilities, we must simultaneously reduce the sunspot distance $L$ in order to raise the decay index at the position of the flux rope apex. The critical decay index for the two configurations is numerically found to lie slightly below $n=1.40$ and 1.80, which respectively correspond to $L=1.2$ and 0.9 chosen here. It should be noted that the stabilizing effect of the guide field is presumably somewhat overemphasized in the analytical equilibrium compared to the conditions in the solar corona because the observations of the shear in flare loops indicate that the guide field decreases strongly already above a few initial flux rope heights \cite[e.g.,][]{SuYN&al2006, Warren&al2011}. The parameters defining this core set of equilibria and of several configurations considered for comparison purposes are compiled in Table~\ref{t:shedding}, together with the (absolute values of the) resulting helicities. 

Figure~\ref{f:helicity} shows that the initial helicity decreases for increasing $\Bet/\Bep$ in Runs~T1--T3. The mutual helicity decreases as the flux rope is more aligned with the ambient flux, and the self helicity decreases as the twist in the flux rope is reduced by the higher external toroidal field (Table~\ref{t:shedding}). Here the mutual helicity contributes only slightly more to the overall decrease than the self helicity. 

The fraction of helicity shed by the eruption decreases as well, which is due to two effects. First, as in Run~T1, almost all of the initial self helicity is shed in Runs~T2 and T3, but self helicity makes up a progressively smaller fraction of the initial helicity in these runs. Second, mutual helicity is shed less effectively, which appears to result from an inability to shed mutual helicity beyond a certain base level (see the rather similar final helicities for T1--T3 in Figure~\ref{f:helicity}). In other words, the system cannot fully approach the potential field. 
This base level of normalized mutual helicity in the configurations T1--T3 is $\approx\!0.06\mbox{--}0.08$, but its value differs for other configurations. It is smaller in Run~K presented below, but is expected to be higher if the legs of the flux rope do not reconnect in the vertical current sheet (as in the 2D standard flare model). In this case, the strongest flux bundles corresponding to the potential field, that is, those connecting the flux rope footpoints with the sources of the strapping field, do not form. 

The complete or nearly complete shedding of the initial self helicity from our configurations does not mean that the post-eruption state does not possess any remaining current density, helicity, or free energy. Currents remain in the two prominent flux bundles that connect the main polarities similarly to the potential field (Figure~\ref{f:current} at $t\!=\!299$). These flux bundles have the same handedness as the original flux rope. However, their ambient flux has built up a weak opposite handedness in a large volume (different from the purely potential ambient field of the original configuration), and so contributes self helicity of opposite sign. Opposing contributions to the total remaining self helicity are found in all of our runs. These cancel out nearly perfectly in Runs~T3 and S and yield a weak self helicity opposite to the initial one in Runs~T1, T2, and K. Their underlying currents permit a significant mutual helicity.

\subsubsection{Kink-unstable case}
\label{sss:kink}

For comparison, we include a kink-unstable run (Run~K in Table~\ref{t:shedding}), whose parameters are similar to the run in \citet{Kliem&al2011}; except for a flatter flux rope shape ($d=0.83$) and a somewhat larger minor radius, $a=0.33$, in that run. A minor fraction of solar eruptions appear to be initiated by the onset of the helical kink instability (see, e.g., the event in \citealt{JiH&al2003} and its modeling in \citealt{Hassanin&Kliem2016} and \citealt{Hassanin&al2016}, which yielded an estimate of $\Phi\approx4\pi$ for the average flux rope twist). As the helical kink saturates already for moderate displacements of the current channel, the subsequent onset of the torus instability is required for evolution into a CME; otherwise a confined eruption is produced. The helical kink lifts the flux rope into the torus-unstable height range in such CMEs; this is also the case in our simulation. For Run~K, we keep the ratio $\Bet/\Bep=0.5$ from Run~T3 and reduce the minor radius $a$ to raise the twist. The sunspot distance $L$ is set to a moderately subcritical value in order to initially stabilize the torus mode but allow its occurrence after the development of the helical kink. Both parts of the initial helicity are further reduced due to the smaller flux content of the thinner flux rope. This run also continues the trends in the  effectiveness of the shedding found for Runs~T1--T3: (1) self helicity, although shed completely, contributes less because it makes up an even smaller fraction of the initial helicity, and (2) the effectiveness of shedding mutual helicity is further reduced. As a result,  the fraction of total helicity shed is only about one-quarter, which is comparable to the value of about one-fifth or less estimated for the configuration in \citet{Kliem&al2011}.

\subsubsection{Effect of flux rope geometry}
\label{sss:geometry}

%~~~~~~~~~~~~~~~~~~~~~~~~~~~~~~~~~~~~~~~~~~~~~~~~~~~~~~~~~~~~~~~~~~~~~~~~~~
\begin{figure}  % [h]                                                Fig. 5
\centering
\resizebox{.95\hsize}{!}{\includegraphics{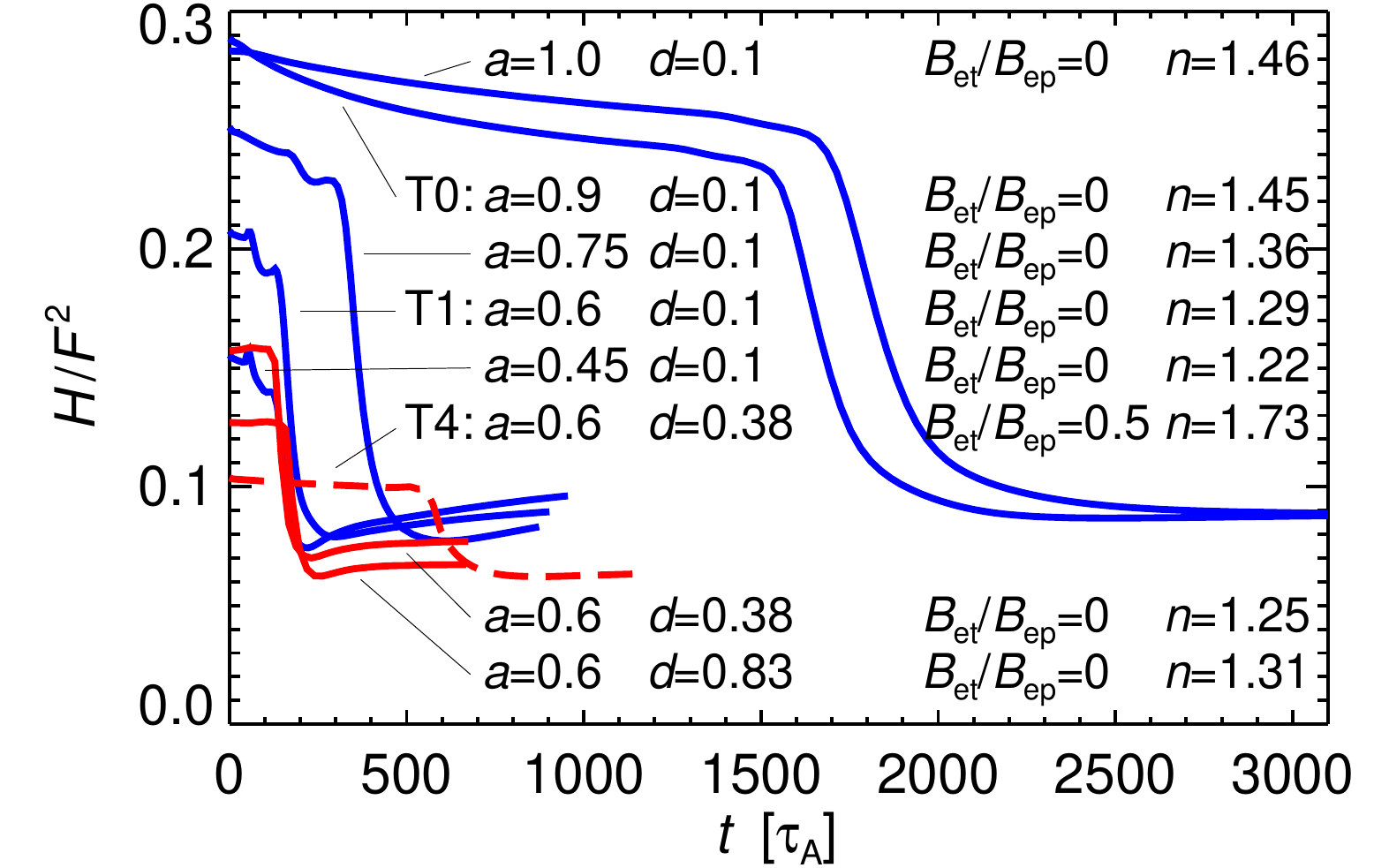}}  % {f5.eps}}
\caption[]{Same as Figure~\ref{f:helicity} for the additional sets of weakly torus-unstable runs with varying minor radius (blue curves) and flatness (red curves) of the initial current channel.} 
\label{f:helicity_ext}
\end{figure}
%~~~~~~~~~~~~~~~~~~~~~~~~~~~~~~~~~~~~~~~~~~~~~~~~~~~~~~~~~~~~~~~~~~~~~~~~~~

To cover a broader range of the conditions on the Sun, we also consider two sets of very weakly torus-unstable equilibria with varying geometric parameters of the flux rope. The kink-unstable configuration of Run~K indicates that the ratio of the flux in the current channel, $F_\mathrm{cc}$, and external poloidal flux, $F_\mathrm{ep}$, which is small in this run due to the small minor radius, may considerably influence both the initial helicity and the conjectured base level of the mutual helicity, and therefore the effectiveness of the shedding. To study these effects, the minor radius is varied in the range $a=0.45\mbox{--}1$. This corresponds to a flux ratio $\mathcal{R}=F_\mathrm{cc}/F_\mathrm{ep}=0.26\mbox{--}1.06$ ($\mathcal{R}=0.42$ in Run~T1). In the second set, the rope is made flatter by setting $d=0.375$ and $d=0.833$, which raises the half-separation of its footpoints from $D_\mathrm{f}=1.1$ in Run~T1 to $D_\mathrm{f}=1.3$ and 1.6, respectively. Both sets of runs use $\Bet/\Bep=0$, and are to be compared to 
Run~T1. 

Figure~\ref{f:helicity_ext} shows a significant increase in the normalized helicity with increasing flux ratio, up to a value of $H_0/F^2=0.29$ for $a=0.9$, which we refer to as Run~T0. The helicity remaining after the ejection of the flux rope shows only a weak increase with $\mathcal{R}$. Consequently, the effectiveness of helicity shedding increases from $\Delta H/F^2=0.084$ (54\%) for $a=0.45$ to $\Delta H/F^2=0.20$ (70\%) for $a=0.9$. As in Run~T1, all of the self helicity but only part of the mutual helicity (17--57\%) is shed. These configurations possess a very weakly varying critical ``sunspot distance'', lying in the range $1.3<L_\mathrm{cr}<1.35$ for $a=0.45\mbox{--}0.9$ and within $1.35<L_\mathrm{cr}<1.4$ for $a=1$. The decay index values $n(z=h_\mathrm{m})$ given in Figure~\ref{f:helicity_ext} correspond to the lower edges of these ranges and vary slightly more than $L$ because the position of the magnetic flux rope axis increases slightly with increasing flux rope thickness (the large-aspect-ratio approximation underlying the expression for the tokamak equilibrium degrades as $a$ increases). 

The highest normalized helicity of a stable configuration in our parametric study is $H_0/F^2=0.27$, found for $a=0.9$ with $d=0.1$ and $L=1.35$. When analyzed with higher spatial resolutions of $\Delta_{x,y,z}=0.02$ and 0.01, this value reduces slightly to $H_0/F^2=0.25$ and appears to converge at this value. (It is worth noting that a decrease in the box size by factors of 2 and 4 in each direction, keeping the resolution at $\Delta_{x,y,z}=0.04$, does not change the computed helicities.) The corresponding values of other relevant parameters are $H_\mathrm{self,0}/H_0=0.315$ and $\mathcal{R}=F_\mathrm{cc}/F_\mathrm{ep}=0.94$. To our knowledge, these are the highest values of $H/F^2$ and $\mathcal{R}$ found so far for stable force-free flux rope equilibria. However, such high-arching and simultaneously very thick flux ropes (filling all space between the flux rope axis and the photosphere) may be rare on the Sun. The observations of extreme ultraviolet (EUV) hot channels indicate smaller cross-sections, typically of the order of $a\sim0.5$, when the channel arches high up in the corona before erupting \cite[e.g.,][]{Reeves&Golub2011, ChengX&al2013, Patsourakos&al2013}. 

Both parts of the normalized helicity decrease as flatter flux ropes are considered. At first sight, this is counter-intuitive because the longer flux rope has a higher twist combined with the same toroidal flux and because the area under the flux rope, through which the external flux $F_\mathrm{ep}$ links with the flux rope, increases with $D_\mathrm{f}$. Both self and mutual helicity indeed increase with increasing $d$, but the flux also increases (because the larger distance to the sources of the strapping field requires a higher source strength in the adopted normalization) and both normalized helicities decrease. This can perhaps be best understood by envisioning a normalization to $R$, that is, a fixed torus with a fixed helicity in the whole space. Increasing $d$ raises the photospheric level toward the upper edge of the torus, so that a progressively smaller fraction of the helicity remains in the coronal half space. The initial partition into self and mutual helicity is the same as in Run~T1, but the critical sunspot distance decreases, and the decay index increases (both do so weakly in the considered range), because the larger footpoint area of the flatter flux ropes increases the stabilizing effect of the line-tying. 

Again, the self helicity is shed completely, indicating the existence of a base level of mutual helicity. This level also decreases weakly from the one in Run~T1, meaning that the effectiveness of helicity shedding decreases moderately from 64\% in Run~T1 to 53\% for the flattest flux rope in our set of equilibria. Within the range of parameters considered, changing the flatness of the flux rope has a weaker influence on the effectiveness of helicity shedding than changing the guide field or the flux content of the rope. 

Finally, in Run~T4, we include a guide field in one of the weakly twisted, flatter flux rope equilibria in order to obtain a more reliable estimate of the minimum helicity shedding by torus-unstable flux ropes. Here we choose $\Bet/\Bep=0.5$ and $d=0.375$ because the equilibrium with $d=0.833$ is completely stable for this guide field strength. The shedding effectiveness of 38\% is found to be only slightly lower than in Run~T3, and the remaining mutual helicity also stays in the range of $\sim\!0.06\mbox{--}0.08$ obtained in Runs~T1--T3 (see Table~\ref{t:shedding} and Figure~\ref{f:helicity_ext}).

\subsection{Model of a solar event}
\label{ss:solar}

To judge the parametric study against the conditions on the Sun, we repeat the modeling of the CME on 2010 April~8 from Active Region NOAA~11060 \citep{Kliem&al2013} with an open upper boundary. This was a weak event (a relatively slow CME with a projected velocity of $\approx\!520$~km\,s$^{-1}$, associated with a weak flare of magnitude B3.7) from a decaying, approximately bipolar active region. Due to the simplicity of the source region, the event should be comparable to our model runs above. It was estimated that the eruption was driven by the torus instability from a threshold value of $n\approx1.5\mbox{--}1.75$, similar to the range of thresholds in Runs~T2 and T3. The modeling of the source region magnetic field indicated a flux rope with a weak to moderate guide field, also similar to runs T2 and T3, and a very small average twist of $\sim\!0.5\pi$. Approximating the external field by the potential field, $\Bet/\Bep\sim0.4$ at the axis of the flux rope in the center of the active region. For this simulation (Run~S in Table~\ref{t:shedding}), we obtain a helicity shedding of $0.42H_0$, similar to Runs~T2--T3 as well. 

The brief increase of the helicity in this run during $t\approx\!100\mbox{--}130$ is a signature of the steepening of the current layer in front of the erupting flux rope (see Figure~5 in \citealt{Hassanin&Kliem2016} for the visualization of a similar configuration). By Lenz's law, the toroidal direction of the current in this layer is opposite to the toroidal flux rope current. However, the toroidal field direction is the same as inside the rope. Therefore, the current layer locally changes the handedness of the field to the opposite of the flux rope's handedness. When this flux leaves the box, the total helicity increases weakly and briefly until the flux rope begins to leave the box. 

It is noteworthy that the normalized initial helicity lies far below the values in our model configurations. This results from the fact that the solar active region contains a large amount of dispersed and rather remote flux that does not contribute to the equilibrium of the flux rope. Such ``exterior flux'' does not exist in the Titov-D\'emoulin equilibrium, where all external flux passes over the polarity inversion line that hosts the flux rope. Exterior flux exists typically in the source regions of solar eruptions, except during the early stage of active-region emergence, which on the other hand does not (yet) launch CMEs. The typical existence of exterior flux can also be seen from the small maximum value of $\approx\!0.022$ for the normalized helicity injected into solar active regions through the photosphere prior to eruptions \citep{LaBonte&al2007}. Our configuration in T1 contains an approximately ten times higher normalized helicity very near the marginal stability state.

\section{Discussion}
\label{s:discuss} 

The ratio $\Bet/\Bep$ of the external toroidal (guide) to the external poloidal (strapping) field is a key parameter in determining self- and mutual helicity of force-free flux rope equilibria and their stability against the torus mode. Here, both $\Bet/\Bep$ and $n$ are varied in ranges that appear to be representative of a large fraction of solar eruptions \cite[e.g.,][]{ChengX&al2020}, and so may be the range of helicity shedding found, $\sim\!(0.4\mbox{--}0.65)H_0$. 

The geometric range covered should be representative as well. Flux ropes forming in filament channels tend to be thick \cite[e.g.,][]{Savcheva&vanBallegooijen2009, SuYN&al2011}, as in our studied range, and flat, flatter than can reasonably be studied using the Titov-D\'emoulin equilibrium. However, the slow rise of filaments or EUV 
hot channels up to the onset height of eruption always yields a considerably arched shape \citep{Nindos&al2015, ChengX&al2020}, with a semicircular shape being not uncommon \cite[e.g.,][]{Schrijver&al2008, ChengX&al2013}. 

The essentially complete shedding of the initial self helicity in our simulation runs is facilitated by the open top boundary. When the top part of the flux rope has left the box completely, the legs can untwist freely. On the Sun, such untwisting is possible if the flux rope reconnects with open ambient flux. However, even if the flux rope remains intact, nearly all of its twist propagates into the interplanetary space. A significant but still minor part of the initial twist and self helicity remain in the corona only if the flux rope legs reconnect with each other \citep{Hassanin&Kliem2016, Hassanin&al2016} or with closed ambient flux. 

The incomplete shedding of mutual helicity in our runs suggests that a base level of mutual helicity may exist that limits the fraction of helicity that can be shed by erupting flux ropes. This level appears to be weakly dependent on the parameters of the initial equilibrium, falling in the range $H_\mathrm{mutual}/F^2\sim0.06\mbox{--}0.08$ for typical source region parameters of solar eruptions. An incomplete shedding of helicity is indeed typically indicated by the considerable shear of the first forming flare loops \citep{SuYN&al2007} and has also been found in simulations of long-term coronal evolution \citep{Mackay&Yeates2012}. Further study is needed to understand its origin. 

Overall, we find that a significant part of the initial relative helicity is shed by a typical flux rope eruption. However, the remaining part is significant as well, on the order of $H_0/2$. Therefore, further mechanisms may be relevant in regulating the helicity budget of the corona (Section~\ref{s:intro}) and merit a quantitative study.

\section{Conclusions}
\label{s:concl} 

Our parametric simulation study of flux rope eruption by ideal MHD instability of a line-tied tokamak equilibrium \citep{Titov&Demoulin1999} demonstrates the helicity shedding conjectured for CMEs. For the most relevant case of torus-unstable flux ropes in bipolar source regions and expected typical partitions of the initial relative helicity into mutual and self helicity in the range $H_\mathrm{mutual,0}/H_\mathrm{self,0}=2.1\dots3.9$ (corresponding to ratios of the equilibrium guide and strapping field components in the range $\Bet/\Bep=0\dots0.5$), we find that a fraction of $\sim\!(0.4\mbox{--}0.65)H_0$ of the initial helicity $H_0$ is shed. A data-constrained simulation of a solar eruption (from Active Region NOAA~11060 on 2010 April~8) yields a shedding of $0.42H_0$, which is consistent with the model results. 

While the initial self helicity (flux rope twist) is completely or nearly completely shed, the effectiveness of shedding mutual helicity decreases with a decreasing initial flux-normalized value of this component. This appears to result from an inability to shed mutual helicity beyond a certain base level, which depends on the parameters of the initial equilibrium and is in the range of $\sim\!0.06\mbox{--}0.08$ for our torus-unstable model configurations. 

The parametric study yields stable equilibria up to a normalized helicity of $H/F^2=0.25$, which, to our knowledge, is higher than what has been found so far \citep{FanY2010}. For this equilibrium, the ratio of the flux in the current channel to the external poloidal flux, $\mathcal{R}=F_\mathrm{cc}/F_\mathrm{ep}=0.94$, is significantly higher than previous estimates of the limiting flux ratio, which lie in the range of $\sim\!0.1\mbox{--}0.5$ \citep{Bobra&al2008, Savcheva&al2012, Kusano&al2020}. The corresponding ratio of self (current-carrying) helicity to total helicity is $H_\mathrm{self}/H=0.315$, which is only slightly higher than a previously suggested limit of $\simeq0.29\pm0.01$ for torus-unstable flux ropes \citep{Zuccarello&al2018}, but lower than a limit of $\sim\!0.45$ for eruptions from newly emerging flux \citep{Pariat&al2017}. 

The flux-normalized helicity in solar data depends strongly on the amount of exterior flux in the magnetogram, that is, flux that is not part of the flux rope equilibrium. Such flux must be separated in order to find a reliable estimate of $H/F^2$ .

\begin{acknowledgements}
We acknowledge the careful reading and constructive comments by an anonymous referee. 
This work was supported by the DFG and by NASA grants 80NSSC17K0016, 80NSSC18K1705, 80NSSC19K0860, 80NSSC19K0082, and 80NSSC20K1274.
\end{acknowledgements}

% \bibliographystyle{aa}
% \bibliography{helshed}

\appendix

\section{Derivation of Eqs. (\ref{e:helicity})--(\ref{e:2D_Poisson})}
\label{appendix:equations}

We start from Eqs. (19--22) in \citet{Valori&al2012}, which can be written as
\begin{equation}
 \vec{A} = \vec{b} + \vec{\hat{z}}\times\int_z^{z_2} \vec{B} \, \mathrm{d}z' \,, \quad
 \vec{A}_\mathrm{C} = \vec{b}_\mathrm{C} + \vec{\hat{z}}\times\int_z^{z_2} \vec{B}_\mathrm{C} \, \mathrm{d}z' \,,
\label{Valori_A+A_C}
 \end{equation}
 where
\begin{equation}
\vec{b}=
\begin{pmatrix}
b_x(x,y)\\
b_y(x,y)\\
0
\end{pmatrix}
\,, \quad
\vec{b}_\mathrm{C}=
\begin{pmatrix}
b_{\mathrm{C},x}(x,y)\\
b_{\mathrm{C},y}(x,y\\
0
\end{pmatrix}
\end{equation}
satisfy
\begin{equation}
\nabla\times\vec{b} = \nabla\times\vec{b}_{\mathrm{C}} = B_z(x,y,z_2) \, \vec{\hat{z}} \,.
\end{equation}
With the choice $z_2=0$ and $\vec{b}=\vec{b}_{\mathrm{C}}$, Eqs. (\ref{Valori_A+A_C}) become
\begin{equation}
 \vec{A} = \vec{b}_\mathrm{C} - \vec{\hat{z}}\times\int_0^{z} \vec{B} \, \mathrm{d}z' \,, \quad
 \vec{A}_\mathrm{C} = \vec{b}_\mathrm{C} - \vec{\hat{z}}\times\int_0^{z} \vec{B}_\mathrm{C} \, \mathrm{d}z' \,,
\label{Valori_A+A_C_special}
\end{equation}
where $\vec{b}_{\mathrm{C}}$ obeys
\begin{equation}
\nabla\times\vec{b}_{\mathrm{C}} = B_z(x,y,0) \, \vec{\hat{z}} \,.
\label{curl_b_C}
\end{equation}
As
\begin{equation}
 \vec{b}_{\mathrm{C}}(x,y) = \vec{A}_{\mathrm{C}}(x,y,0) \,,
\end{equation}
the first equation in Eqs.~(\ref{Valori_A+A_C_special}) is identical to Eq. (\ref{e:vector_potential_A}).

As suggested by \citeauthor{Valori&al2012}, we choose $\vec{b}_{\mathrm{C}}$ solenoidal, which is possible because both the curl (given by Eq. (\ref{curl_b_C})) and the divergence of $\vec{b}_{\mathrm{C}}$ can be prescribed. Therefore, $\vec{b}_{\mathrm{C}}$ can be represented in the form 
\begin{equation}
\vec{b}_{\mathrm{C}} =
\begin{pmatrix}
- \partial_y g_{\mathrm{C}} \\
\partial_x g_{\mathrm{C}} \\
0
\end{pmatrix}
= \vec{\hat{z}}\times\nabla g_{\mathrm{C}}
\label{stream_function}
\end{equation}
by means of a stream function
$g_{\mathrm{C}}(x,y)$.
The second equation 
in Eqs.~(\ref{Valori_A+A_C_special}) then becomes
\begin{equation}
 \vec{A}_{\mathrm{C}} = \vec{\hat{z}}\times\nabla g_{\mathrm{C}} + \int_0^z \nabla\phi_{\mathrm{C}} \, \mathrm{d}z'
 =\vec{\hat{z}}\times\nabla \left( g_{\mathrm{C}} + \int_0^z \phi_{\mathrm{C}} \, \mathrm{d}z' \right) \,,
 \label{vector_potential_A_C-Appendix}
\end{equation}
which is identical to Eq. (\ref{e:vector_potential_A_C}), while
Eq. (\ref{curl_b_C}) takes the form of the Poisson equation
\begin{equation}
 \left(\partial_x^2 + \partial_y^2\right) g_{\mathrm{C}}(x,y) = B_z(x,y,0) \,,
\end{equation}
which is identical to Eq. (\ref{e:2D_Poisson}).

The relative magnetic helicity in the box volume $V$ is calculated using the formula of \citet{Finn&Antonsen1985}:
\begin{align}
 H=&\int_V \left(\vec{A}+\vec{A}_\mathrm{C}\right)\cdot\left(\vec{B}-\vec{B}_\mathrm{C}\right)\, \mathrm{d}^3\vec{r} 
=\int_V \vec{A}\cdot\vec{B}\, \mathrm{d}^3\vec{r} \nonumber\\
&-\int_V \vec{A}_\mathrm{C}\cdot\vec{B}_\mathrm{C}\, \mathrm{d}^3\vec{r}
+\int_V \left(\vec{A}_\mathrm{C}\cdot\vec{B}-\vec{A}\cdot\vec{B}_\mathrm{C}\right)\, \mathrm{d}^3\vec{r.}
\label{H_FA}
\end{align}
For the first integral in the second line
of Eq.~(\ref{H_FA}) we have
\begin{align}
\int_V \vec{A}_\mathrm{C}\cdot\vec{B}_\mathrm{C}\, \mathrm{d}^3\vec{r}
&= -\int_V \nabla\phi_\mathrm{C}\cdot\vec{A}_\mathrm{C}\, \mathrm{d}^3\vec{r} \nonumber\\
&= -\int_V\nabla\cdot\left(\phi_\mathrm{C}\vec{A}_\mathrm{C}\right)\, \mathrm{d}^3\vec{r} \nonumber\\
&= -\int_{\partial V} \phi_\mathrm{C}\vec{A}_\mathrm{C}\cdot\mathrm{d}\vec{S} \nonumber\\
&= -\int_{\text{Side faces}} \phi_\mathrm{C}\vec{A}_\mathrm{C}\cdot\mathrm{d}\vec{S} \,,
\label{int_A_C_B_C}
\end{align}
where we make use of $\nabla\cdot\vec{A}_\mathrm{C}=0$  and $\vec{A}_\mathrm{C}\cdot\vec{\hat{z}}=0$.
The second integral in the second line
of Eq.~(\ref{H_FA}) can finally be rewritten as
\begin{align}
 \int_V \left(\vec{A}_\mathrm{C}\cdot\vec{B}-\vec{A}\cdot\vec{B}_\mathrm{C}\right)\, \mathrm{d}^3\vec{r}
&= \int_V \nabla\cdot\left(\vec{A}\times\vec{A}_\mathrm{C}\right)\, \mathrm{d}^3\vec{r}
\nonumber\\
&= \int_{\partial V} \left(\vec{A}\times\vec{A}_\mathrm{C}\right)\cdot\mathrm{d}\vec{S} \nonumber\\
&=\int_{\text{Top face}} \left(\vec{A}\times\vec{A}_\mathrm{C}\right)\cdot\mathrm{d}\vec{S} \,,
\label{int_H_FA_mixed}
\end{align}
where we take advantage of the fact that $\vec{A}\times\vec{A}_\mathrm{C}$ is purely vertical (parallel to the $z$ axis) and $\vec{A}(x,y,0)=\vec{A}_\mathrm{C}(x,y,0)$.
Inserting Eqs. (\ref{int_A_C_B_C}) and (\ref{int_H_FA_mixed}) into Eq. (\ref{H_FA}) leads to Eq. (\ref{e:helicity}).

\end{document}